# Displacement calibration of optical tweezers with absolute gravitational acceleration

**JIANYU YANG[1], NAN LI[1,4], XUNMIN ZHU[2], MING CHEN[1], XINGFAN CHEN[1], CHENG LIU[1], JIAN ZHUANG[3], AND HUIZHU HU[1,2,5]**

[1]*State Key Laboratory of Modern Optical Instrumentation, College of Optical Science and Engineering, Zhejiang University, Hangzhou 310027, China*
[2]*Quantum Sensing Center, Zhejiang Lab, Hangzhou 310000, China*
[3]*HIWING Technology Academy of CASIC, Beijing, 100074, China*
[4]*nanli@zju.edu.cn*
[5]*huhuizhu2000@zju.edu.cn*

**Abstract:** In recent years, levitated particles of optical traps in vacuum have shown enormous potential in precision sensor development and searching for new physics. The accuracy of the calibration relating the detected signal to absolute displacement of the trapped particle is a critical factor for absolute measurement performance. In this paper, we suggest and experimentally demonstrate a novel calibration method for optical tweezers based on free-falling particles in vacuum, where the gravitational acceleration is introduced as an absolute reference. Our work provides a calibration protocol with great certainty and traceability, which is significant in improving the accuracy of precision sensing based on optically levitated particles.

## 1. Introduction

Levitated particles in optical tweezers could ultimately be considered as an extremely sensitive sensor due to the ideal isolation from the environment [1-6]. Excellent force [7-11] and acceleration sensitivity [9, 12] had been achieved in recent years. Moreover, these systems also paves the way for the investigation of new fundamental interactions [13-17] and macroscopic quantum physics in unexplored scale of large masses [18-21].

Any fluctuation acting on the levitated particle gives rise to a displacement, which can be recorded by scattered optical signal. Thus the calibration between the detected signal and the absolute displacement of the trapped particles is of great significance, especially for those measurements of absolute physical quantities [7]. The most commonly used calibration methods are passive calibration methods based on the power spectrum density calculated from the particle's Brownian trajectory and invoke equipartition of the potential or kinetic energy amongst all degrees of freedom [22-24]. These calibration methods rely on a key assumption that detector signal is linear in the particle displacement while nonlinear detector response is not applicable [24]. Also, the uncertainty of trapped particle's mass and bath temperature will bring unknown influence to the calibration precision. In addition, under special condition, for instance orbital revolution mode, the trapped particle does not follow a simple motion mode. In this case, calibration methods based on power spectrum density will not able to work correctly [25].

In addition to passive calibration methods based on power spectrum density, there are also active calibration methods such as placing electrodes around the optical trap and apply a modulating force on the particle [9, 10, 12]. These methods are suitable for nonlinear conversion factor. However, these methods do not have an absolute reference. The uncertainty of electrode parallelism and spacing will lead to the inaccuracy of the calculation of electric field and affect the calibration. Also, the uncertainty of the particle mass will affect the calibration, too. In addition, the active calibration methods often need to build complex calibration devices around the optical trap, which increases the complexity and assembly difficulty of the system [9, 10, 12].

In this paper, we propose and experimentally demonstrate a novel calibration scheme with free-falling particles which use gravitational acceleration as an absolute reference. This method does not require complex calibration devices and can avoid the influence of particle mass and bath temperature uncertainty on the calibration accuracy and calibrate the relationship between the detected signal and displacement of particle even if the relationship is nonlinear or the trapped particle is in a complicated motion mode.

## 2. Theory

The calibration scheme we proposed is based on a free-falling and oscillation particle. The calibration procedure is illustrated in Fig. 1. First, we capture a microsphere particle in optical trap in vacuum. Then, we turn off the trapping beams, so the microsphere will fall due to the gravity. After a falling time of $t_1$, we turn on the trapping beams again, and the microsphere will be recaptured as long as the potential well depth is larger than 10kBT [1] (T is the Center of Mass temperature of the particle). Because of the deviation from the equilibrium position acquired during free-falling process, the microsphere obtains potential energy of the optical trap, which results in an underdamped oscillation at the resonance frequency. The amplitude of the oscillation will decrease exponentially over time due to the damping of surrounding molecular, which will be recorded by trapping/detection beam. After time $t_2$, the oscillation amplitude becomes negligible compare to the initial amplitude. Then we turn off the trapping beam again and repeat the release and recapture cycle constantly. We record all periods of oscillation signal for further analysis.

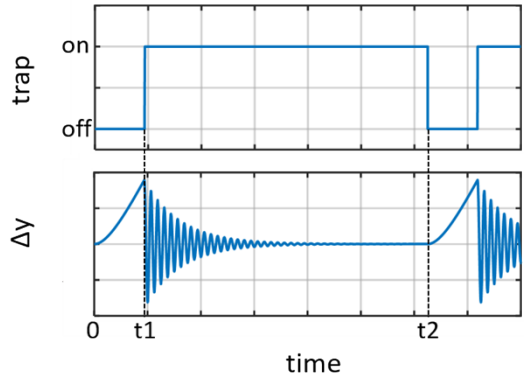

Fig. 1. The principle of calibration. The microsphere is optically trapped and then experiences a free-falling process when we switch the trap off at t=0. At t=t1, the trap is switch back on; the motion of microsphere is then transformed into underdamped oscillation. At t=t2, when the amplitude becomes negligible due to the damping of surrounding molecule, we turn off the laser beam again and repeat the falling-oscillation cycle periodically.

In order to display the physical process of the falling-oscillation motion of the microsphere more intuitively, we first ignore the Brownian motion of the microspheres. The vertical displacement of the particle can be described by the equation

$$\frac{\partial^2 y}{\partial t^2} + \Gamma_0 \frac{dy}{dt} + \Omega^2 y f(t) = g,$$

$$f(t) = \begin{cases} 0, nT < t < nT + t_1 \\ 1, nT + t_1 < t < (n+1)T \end{cases} \quad (1)$$

where $\Gamma_0$ is damping rate of the surrounding molecule, $\Omega$ is resonance angular frequency of the trap, g is the acceleration of gravity, and $T = t_1 + t_2$ is the time of one falling-oscillation period. Solve this equation in time domain, we obtain

$$y = \frac{g}{\Gamma_0} t + \frac{1}{\Gamma_0}(\frac{g}{\Gamma_0} - v_0)\left[\exp(-\Gamma_0 t - 1) + y_0\right], f(t) = 0 \quad (2)$$

and

$$y = A\exp(\alpha t)\sin(\beta t + \varphi) + \frac{g}{\Omega^2}, f(t) = 1 \quad (3)$$

where $v_0$, $y_0$ is the final velocity and final displacement of the previous period, the underdamped oscillation parameter $\alpha = -\frac{\Gamma_0}{2}$, $\beta = \sqrt{\Omega^2 - \left(\frac{\Gamma_0}{2}\right)^2}$, $A = \sqrt{c_{fall}^2 + \left[\frac{1}{\beta}(v_{fall} - \alpha c_{fall})\right]^2}$, $\varphi = \sin^{-1}\frac{c_{fall}}{A}$, and $c_{fall}$ is the falling distance, $v_{fall}$ is the final velocity of the falling. The solution of the equations shows that every falling-oscillation period depends on the final state of the previous period. The trapped particles obtain energy during falling and consume it during oscillation because of the damping of surrounding molecules. We use the kinetic energy after falling to describe the energy evolution in each period. From the simulation results in Fig. 2(a), we can see there are two kinds of falling-oscillation motion mode. When the energy obtained from falling is larger than the energy consumed during oscillation, the energy and amplitude of the oscillation will increase exponentially, and soon the microsphere cannot be recapture after falling and escape from the trap (red line). On the other way, if the energy obtained from falling can be consumed completely during the oscillation, for example when the damping is more significant than the divergent case, the motion of the microsphere will converge rapidly as long as the microsphere can be recapture after falling (green line). The difference between the final state of adjacent periods will decrease exponentially, and every falling-oscillation period will follow a static motion mode. A typical static falling-oscillation period is shown in Fig. 2(b). Which falling-oscillation mode the microsphere follows depends on the ratio of $t_1$ to $T$ and the pressure. For a falling time of 500 μs and the ratio of 0.8%, the critical pressure of the two mode is about 1 mbar.

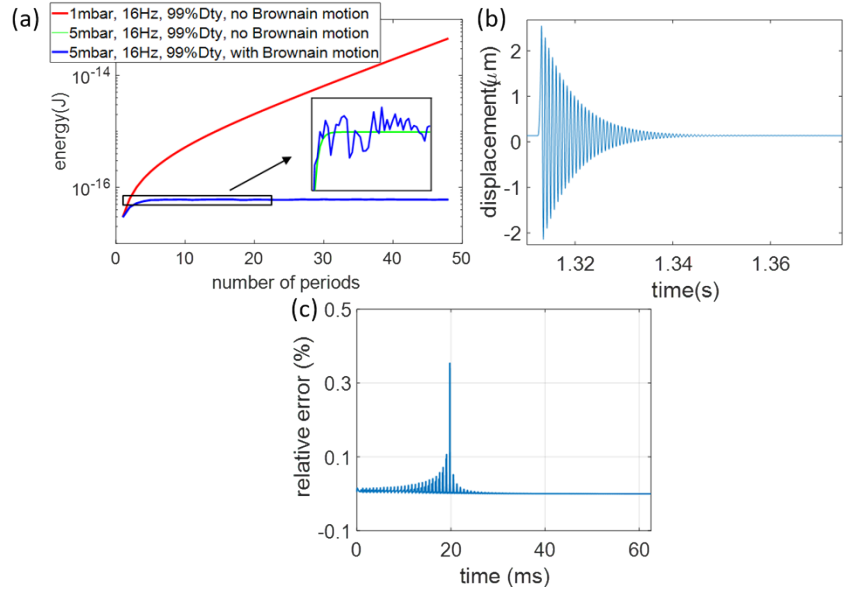

Fig. 2. (a) Energy of three motion mode. When the energy obtained from falling is larger than the energy consumed during oscillation, the energy is divergent (red). When the obtained energy is all consumed during oscillation, the energy is convergent (green). When considering Brownian motion of the particle under the same conditions as the green line, the energy of each period will fluctuate but is also convergent (blue). (b) A typical convergent falling-oscillation period with a falling time of 500 μs, the ratio of falling time to period is 0.8% and the pressure is 1 mbar. (c) Relative error between the multi-period average signal (about 2000 periods) and the static signal without considering Brownian motion. The relative error is relatively large near the zero points, but the maximum error is less than 0.4%.

However, if we consider the Brownian motion of the microsphere, things will be different [23]. For a previous convergent motion mode, instead of being consistent, the energy of each period will fluctuate because of Brownian motion as observed in Fig. 2(a) (blue line). However, when we average each period (around 2000 periods) and subtract it from the static period without considering Brownian motion in Fig. 2(c), we discover that the difference, which is relatively bigger when the displacements are close to zero, between the two situation is negligible. Therefore, we can eliminate the influence of Brownian motion by means of multi-period averaging.

We assume a third order detection nonlinearity and the relation between the detection signal and the displacement is

$$V(t) = a_1 y^3(t) + a_2 y(t) + \xi_V(t) + V_{offset} \qquad (4)$$

where the $\xi_V(t)$ is the white noise of detection, $V_{offset}$ is the offset signal. The influence of the white noise can be eliminated by means of multi-period averaging. The averaged signal after multi-period averaging is

$$\overline{V}(t) = a_1 \overline{y^3(t)} + a_2 y_{sta}(t) + V_{offset} \qquad (5)$$

where $y_{sta}(t)$ is the static falling-oscillation period. Using the same simulation method in Fig. 2(c), we can also obtain $\overline{y^3(t)} = y_{sta}^3(t)$, so the equation becomes

$$\overline{V}(t) = a_1 y_{sta}^3(t) + a_2 y_{sta}(t) + V_{offset} \qquad (6)$$

The equation also applies to higher order nonlinear system when the difference between $\overline{y^n}$ and $(y_{sta})^n$ is within the acceptable margin of error.

So, using the multi-period averaging method we can eliminate the influence of Brownian motion and the noise. That is, we can take the static falling-oscillation period obtained without considering Brownian motion as the standard displacement corresponding to the multi-period averaging signal.

## 3. Experimental setup

A schematic of the experimental setup is shown in Fig. 3. We trap a microsphere with a diameter of 10 μm in our dual-beam optical trap at the pressure of 5 mbar. The optical trap is created by two weakly focused 1064 nm counter-propagating (CP) orthogonally polarized beams with roughly equal power. The power of each trapping beam is 200 mW and the numerical aperture is 0.02. The CP beams are 15 μm foci offset in axial direction to create a more stable trap [10], and one of them is utilized for three axes position detection after transmitted from the particle, and transform into detection system. We use a D-shape mirror position detection system to detect the motion of the microsphere in X, Y and Z directions [25]. The detection signals are collected by data acquisition cards (DAQ cards) with a sampling rate of 250 kHz. We use a square-wave modulated acousto-optic modulator (AOM) as an optical trap switch.

As analyzed before, the oscillation time should be long enough to consume the energy which microsphere obtained during the fall. The falling time should also be chosen properly. If the falling time is too long, the displacement will exceed the trap or the detection range. Nevertheless, if the falling time is too short, the sampling points of large displacement at the beginning of damping stage will be too few and reduce the calibration accuracy. Based on considerations mentioned above, we chose square waves with frequency of 16 Hz and duty cycle ranging from 98.95% to 99.10% to modulate AOM in our experiment, corresponding to a falling time ranging from 590 μs to 690 μs.

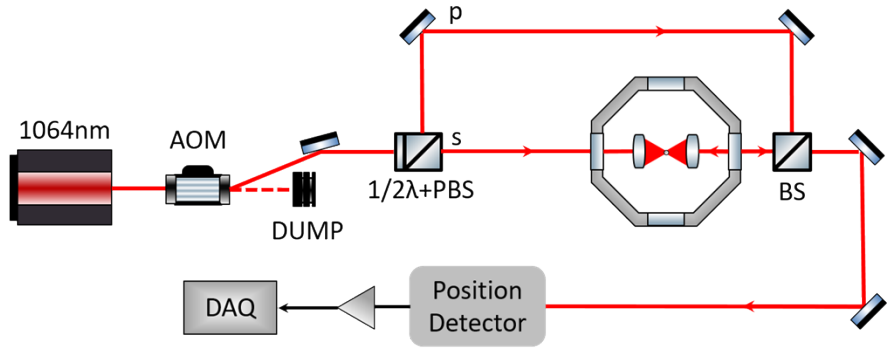

Fig. 3. Experimental setup. A microsphere with a diameter of 10 μm is optically levitated in our dual-beam optical trap created by two weakly focused 1064 nm, 200mW counter-propagating orthogonally polarized beams at the pressure of 5 mbar. The numerical aperture of trapping beam is 0.02. The AOM works as an optical switch to modulate the laser power. The motion of the levitated particle is detected by a position detection module and the detection signals are collected by DAQs.

## 4. Calibration

Fig. 4(a) shows a typical detection signal of one falling-oscillation period with a free-falling time of 656 μs. The signal is seriously affected by Brownian motion and noise. Fig. 4(b) shows the signal processed by multi-period averaging. After multi-period averaging, the signal presents an obvious underdamped oscillation form, which proves the effectiveness of multi-period averaging method. To measure the conversion coefficient $a_1$ and $a_2$, we need to calculate the static displacement corresponding to the averaged signal. To do this, we fit the averaged signal to equation 3 and equation 6 to obtain the resonance angular frequency $\Omega$ and the damping rate $\Gamma_0$ of optical trap, which are equivalent to $\alpha$ and $\beta$ in equation 3. Because the frequency with highest intensity in the spectrum of the signal is independent of nonlinearity, we can easily obtain the angular frequency $\beta$ of the underdamped motion by taking a Fourier transform of the signal. The fitting $\beta$ of signal in Fig. 4(b) is 1374.19×2π Hz.

However, it is impossible to obtain $\alpha$ by fitting the equations if we consider the nonlinearity because the number of variables exceeds the fitting threshold. Whereas, when the displacement of the microsphere is short enough, the nonlinearity is negligible. The fitting equation can be simplified as

$$V = A`\exp(\alpha t)\sin(\beta t+\varphi)+V_{offset} \tag{7}$$

with which we can fit $\alpha$. We use a set of simulation signals to verify the correctness of the method. The signals assume a conversion relationship of $a_1 = 5\times10^{15}$ V/m$^3$, and $a_2 = 4.5\times10^5$ V/m, resonant angular frequency $\Omega$=1340×2π Hz, and damping rate $\Gamma_0 = 250$ Hz, corresponding to $\alpha = -125$ Hz. We fit the signals with different falling time using equation 7. Fig. 4(c) shows the relationship between fitting $\alpha$ and the falling time. As can be seen from the result, when the falling time is relatively short, the fitting $\alpha$ is close to the true value and does not change with the falling time. However, when the falling time is too long, the fitting $\alpha$ is no longer a constant because the nonlinearity of the conversation relationship cannot be neglected and the fitting results are wrong. Since the damping rate of the trap is a constant, the correct fitting results obtained from short falling time is also correct in cases of long falling time. Fig. 4(d) shows the relationship between fitting $\alpha$ and falling time in our experiment. We can see the fitting result starts to go wrong when the falling time is 688 μs. From the result we obtain $\alpha = -132.58$ Hz, corresponding to trapping parameters $\Omega$=1373.3×2π Hz and $\Gamma_0 = 265.15$ Hz.

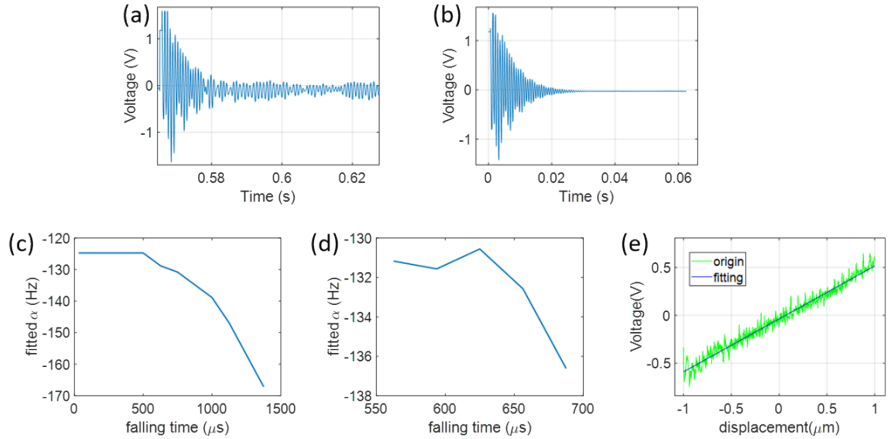

Fig. 4. (a) a typical detection signal of one falling-oscillation period. (b) the multi-period averaged signal, which shows an obvious underdamped oscillation form. (c) The results of fitting coefficient $\alpha$ of simulation nonlinear detection signals at different falling time without considering nonlinear conversion relationship. The fitting results are close to the true value when the falling time is relatively short. (d) The results of fitting coefficient $\alpha$ of experimental signals at different falling time without considering nonlinear conversion relationship. (e) The relationship between the averaged experimental signal and the corresponding static

displacement with a falling time of 656 μs (green) and a third order odd nonlinear fitting to the signal (blue). The fitted third-order coefficient is 0.8521 V/m³ and the fitted first-order coefficient is 5.533×10⁵ V/m.

With $\Omega$ and $\Gamma_0$, now we have all the parameter needed to calibrate the conversion relationship between signal voltage and displacement of trapped microsphere. As discussed in equation 6, every simple point in averaged voltage signal corresponds to a displacement in static motion mode. According to equation 2 and equation 3, we can calculate the static displacement using the fitted $\Omega$, $\Gamma_0$, and the gravitational acceleration, which introduces an absolute reference for calibration. Fig. 4(e) shows the relationship between the averaged experimental signal and the corresponding static displacement with a falling time of 656 μs (green line). We arrange the signal by displacement to get a better view of the relationship between them. Our program fine-tunes the phase of displacement to obtain a more accurate correspondence. By the nonlinear fitting of the relationship between the voltage and the displacement, we obtain $a_1 = 0.8521$ V/m³, and $a_2 = 5.533 \times 10^5$ V/m. From the result we can see the nonlinearity of the conversion relationship in our experiment is negligible. Thus, we verified that the conversion relationship between signal and displacement is linear in our experimental range with a conversion coefficient of $[5.533 \pm 0.206(\text{stat.}) \pm 0.023(\text{sys.})] \times 10^5$ V/m. The system uncertainty is generated by AOM switching time. As a comparison, the conversion coefficient obtained by fitting power spectrum is $1.450 \times 10^6$ V/m [22]. Fig. 5 shows the conversion coefficients calibrated by different falling time in one experiment with a same levitated microsphere. The standard deviation of the results is 4.00%, and the maximum deviation of the results is 5.86%. Therefore, our method can verify and calibrate the nonlinearity of the conversion relationship between the detection signal and the displacement of trapped microsphere, which introduce the gravitational acceleration as an absolute reference for calibration to obtain a higher certainty.

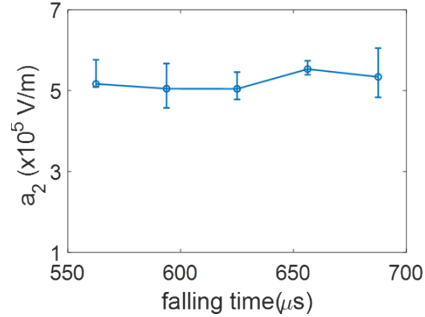

Fig. 5. The conversion coefficients calibrated by different falling time in one experiment with a same levitated microsphere. The standard deviation of the results is 4.00%, and the maximum deviation of the results is 5.86%.

## 5. Discussion

The parameters required by our calibration method are falling time, oscillation time, time interval between sampling points, resonance angular frequency and damping rate of the trap, and the local gravitational acceleration. The falling time and oscillation time are determined by frequency and duty cycle of the square wave which modulates the AOM. The time interval between sampling points is determined by the sampling rate of the DAQ card. The resonance angular frequency and damping rate of the trap can be accurately obtained by fitting the averaged signal. Finally, the local gravitational acceleration can be obtained with great precision in many ways, which introduce an absolute reference for our calibration. Compared with calibration methods based on the power spectrum of microsphere's motion, our method is not affected by the uncertainty of temperature and microsphere mass [24]. At the same time, the proposed method modulates the motion of the microsphere by gravity, and introduce the gravitational acceleration as an absolute reference. Therefore, compare with other active calibration method using other modulation force like electric force, the proposed method is more traceable. What's more, our method only relies on one AOM installed in the optical path, which has been installed in many systems for modulation of laser power, and does not require additional devices around the optical trap [9, 10, 12], which makes our method simple to implement and reduce the complexity of the active calibration system. Our method can also calibrate the nonlinearity of the detection system, which is negligible in our system but becomes significant in the cases of larger microspheres trapped by magnetic traps or trapped microspheres with larger range of displacement.

## 6. Conclusion

We have demonstrated a calibration method for detection systems in optical tweezers systems using free-falling particles. Our method introduces gravitational acceleration as an absolute reference and can calibrate the nonlinearity of the detection systems. What's more, the proposed method is also very simple to implement compare with other active calibration methods.

Our method can help improve the accuracy of precision measurement systems based on levitated particles, which has great application value in the measurements of Casimir force, non-Newtonian gravity, as well as sensitive electromagnetic sensing. This calibration method can also find applications in magnetic traps detection system, whose linearity is more significant.

**Funding.** National Natural Science Foundation of China (No. 62075193), Zhejiang Provincial Natural Science Foundation of China under Grant No. LD22F050002, Major Scientific Research Project of Zhejiang Lab, China (No. 2019MB0AD01) and National Program for Special Support of Top-Notch Young Professionals, China (No. W02070390).

**Disclosures.** The authors declare no conflicts of interest.

**Data availability.** Data underlying the results presented in this paper are available from the corresponding author upon reasonable request.